\documentclass{emulateapj}

\slugcomment{Submitted to ApJL}

\shorttitle{Far-UV sensitivity of COS}
\shortauthors{McCandliss et al.}

\begin{document}


\title{Far-UV sensitivity of the Cosmic Origins Spectrograph}


\author{Stephan R. McCandliss}
\affil{Department of Physics and Astronomy, The Johns Hopkins University, Baltimore, MD  21218}
\email{stephan@pha.jhu.edu}
\author{Kevin France, Steven Osterman, and James C. Green }
\affil{Center for Astrophysics and Space Astronomy,
University of Colorado, 
Boulder, CO 80309-0389}
\author{Jason B. McPhate }
\affil{University of California, Berkeley
Space Sciences Lab 7 Gauss Way
Berkeley, CA 94720-7450}
\author{ Erik Wilkinson }
\affil{Southwest Research Institute, 
1050 Walnut St, Suite 300 
Boulder, CO 80302}




\begin{abstract}
We demonstrate that the G140L segment B channel of the Cosmic Origins Spectrograph (COS) recently installed on the {\it Hubble Space Telescope (HST)} has an effective area consistent with $\sim$ 10 cm$^2$ in the bandpass between the Lyman edge at 912 \AA\ and Lyman~$\beta$, rising to a peak in excess of 1000 cm$^2$ longward of 1130 \AA. This is a new wavelength regime for {\it HST} and will allow opportunities for unique science investigations.  In particular, investigations seeking to quantify the escape fraction of Lyman continuum photons from galaxies at low redshift, determine the scale-length of the hardness variation in the metagalactic ionizing background over the redshift range 2 $< z \lesssim$ 2.8, measure the ratio of CO to H$_2$ in dense interstellar environments with $A_V >$ 3, or harness the high temperature diagnostic power of the \ion{O}{6} $\lambda\lambda$ 1032, 1038 doublet can now be carried out with unprecedented sensitivity.
\end{abstract}


\keywords{instrumentation: spectrographs,  ultraviolet: general}



\section{Introduction}

From the inception of the Cosmic Origins Spectrograph (COS) \citep{Morse:1998,Green:2003} as an instrument for {\it Hubble Space Telescope (HST)} it was appreciated that the combination of a single bounce reflection grating with the large geometric collecting area ($\sim$ 40,000 cm$^2$), and the $\sim$ 15 \% reflectivity at 1000 \AA\ of the Magnesium Fluoride over-coated Aluminum (MgF$_2$/Al) mirrors \citep{Hunter:1971}, could provide useful sensitivity in the far-UV bandpass between the rather sharp MgF$_2$/Al discontinuity at $\sim$ 1150 \AA\ and the Lyman edge at 912 \AA.  The possibility that COS could work below 1150 \AA\ was intriguing because of the potential for new science.  A low spectral resolution far-UV channel could be used for new studies of the steeply increasing populations of faint galaxies and quasars not detectable with the higher spectral resolution channels offered by the Far-Ultraviolet Spectroscopic Explorer {\it FUSE}.  Moreover, {\it FUSE}
is no longer operational and as such the community has lost its window on a set of critical spectral diagnostics, such as for H$_2$ and \ion{O}{6}, that are only available below 1150 \AA.

The COS instrument team found that the windowless detector design and the low resolution G140L grating $R (\equiv \Delta\lambda/\lambda) \sim$ 2000, lent itself easily to the inclusion of a non-optimized far-UV channel in detector Segment B, spanning $\approx$ 100 - 1100 \AA. Its expected effective area was anticipated to be about 100 - 200 times lower \citep{Froning:2009} than that of COS longward of 1150 \AA.  However, this estimation was uncertain.  It was known that upon return to earth the WFPC-1 pickoff mirror had been severely contaminated during its time on orbit with a normal incidence reflection of $\approx$ 75\% at 1500 \AA, falling to $\sim$ 1 \% at 1216 \citep{Tveekrem:1996}.  The use of the G140L segment B was unsupported for the {\it HST} Cycle 17 round of proposals, because of the unknown level of contamination that the {\it HST} Optical Telescope Assembly (OTA) had suffered in the nearly 20 years since its launch. 

Here we present the first evidence that any OTA mirror contamination in the far-UV is inconsequential and that in fact the sensitivity below 1000 \AA\ rivals that of the Silicon Carbide channels on {\it FUSE} \citep{Dixon:2009}.  We discuss some new science opportunities enabled by this new capability and conclude that "Lyman limited" instrumentation is a fesible option for future large aperture UV telescopes.

\section{Observations}

On 04 September 2009 the hot DA white dwarf WD 0320-539 \citep[GSC 08493-00891, T$_{eff}$ = 32.9 K, $\log g$ = 7.89, $V$ = 14.9][]{Vennes:2006} was observed in the G140L channel of COS as part of an external detector flat-field program, number 11491.  The data were acquired in 3 focal plane split (FP/Split) observations, each 640 seconds long, yielding a total on target time of 1920 seconds.   Standard pipeline processing \citep{Kaiser:2008} was used to combine the two-dimensional images of the three separate FP/Split spectra into a single contiguous spectrum, the so called \verb+"*_corrtag_b.fits"+ image, from which we performed a manual extraction and analysis.  We report here only on the Segment B spectrum.  This object has also been observed several times by {\it FUSE} at a resolving power  $R \sim$ 20,000.  The spectrum, provided through MAST, incorporates the final {\it FUSE} calibration \citep{Dixon:2009} and will serve as our absolute calibration reference. The data files for the COS and FUSE spectra used in this work are listed in Table~\ref{tab1}.

\section{Analysis}

The spectra for WD 0320-539 were extracted from the two-dimensional fits files using a rectangular aperture 50  pixels high.  Counts with pulse heights $<$ 3 and $>$ 29 were eliminated.  A detector background region, located 100 pixel below the object region, was similarly extracted and subtracted from the object.    The G140L Segment B wavelength scale is 0.08 \AA\ pixel $^{-1}$ and there are approximately 6 pixels per resolution element.  Further information on the COS instrument can be found in an ``in preparation'' paper by Green et al., or by consulting the COS instrument handbook maintained by the Space Telescope Science Institute.  

\begin{figure}
\includegraphics[angle=90,scale=.65]{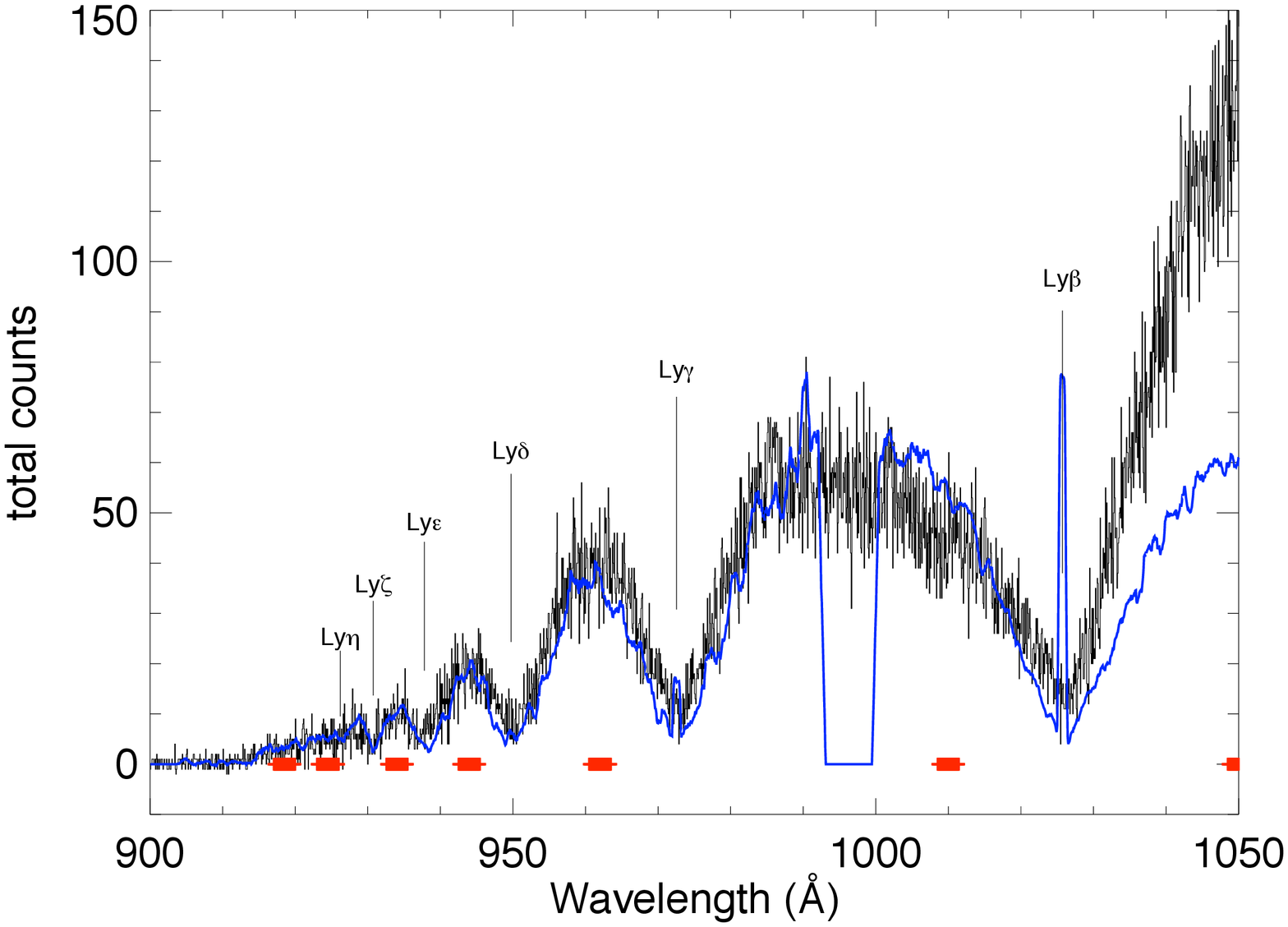}
\caption{The total counts as a function of wavelength acquired from the hot white dwarf WD 0320-539 in 1920 s by the COS G140L Segment B  are shown in black.  The blue line is the {\it FUSE} spectrum of the same object multiplied by a constant factor, smoothed by 100 pixels and resampled to the linear wavelength scale of the G140L Segment B.  The red points at the bottom of the plot mark the continuum fitting regions. The gap in the {\it FUSE} spectrum is due to non overlapping detector segements.  See text for details. \label{fig1}}
\end{figure}

In Figure~\ref{fig1} we show the total count spectrum on a linear scale plotted from 900 -- 1050 \AA, using a linear wavelength solution.  The progression of the pressure broadened Lyman series characteristic of a high gravity hot white dwarf is evident.  The {\it FUSE} spectrum, smoothed by 100 pixels, rebinned to the G140L Segment B wavelength grid and scaled by a constant factor of 8 $\times$ 10$^{13}$ counts (ergs cm$^{-2}$ s$^{-1}$ \AA$^{-1}$)$^{-1}$, is overplotted in blue.  The count spectrum can be faithfully converted to the stellar flux for WD 0320-539 over most of the displayed bandpass using this constant factor, demonstrating the effective area is also nearly constant. There are some slight mismatches in the line shapes, but overall the agreement is excellent.

\begin{figure} 
\includegraphics[angle=90,scale=.65]{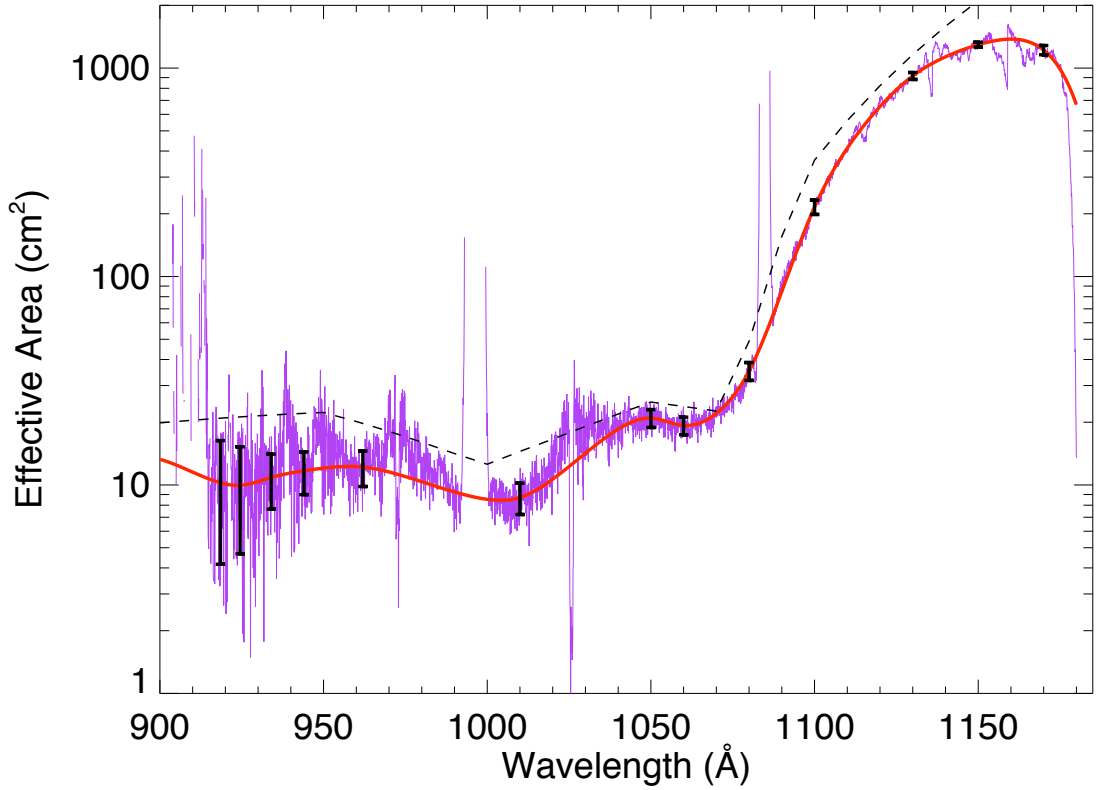}
\caption{Effective area for G140L Segment B.  The purple line is the result of dividing the COS spectrum by a {\it FUSE} spectrum of the same object.  The red line is a spline fit to the result.  The $\pm$1 $\sigma$ error bars for the thirteen 3 \AA\ wide wavelength average regions used for the spline fit are shown in black.  The dashed line is a model for the effective area based on measured detector, grating and mirror efficiencies.  The G140L Segment B effective area $\approx$ 0.6 times the model, see the text for details.   The discontinuities at the long wavelength end of the COS spectra are fixed patterns caused by detector high voltage grids. \label{fig2}}
\end{figure}

The effective area curve, shown on a semi-logarithmic scale in Figure~\ref{fig2}, was derived simply by dividing the {\it FUSE} spectrum into the COS count spectrum and applying the appropriate time ($t$ = 1920 s), energy ($hc/\lambda$) and dispersion ($d\lambda$ = 0.08 \AA) scalings.  A spline fit based on 13 points averaged over 3 \AA\ intervals at wavelengths chosen to avoid the Lyman series line cores  ($\lambda $ = [918.5, 924.5, 934, 944, 962, 1010, 1050, 1060, 1080, 1100, 1130, 1150, 1170] was employed to estimate the effective area.  Points from the spline fit and associated errors, sampled on 20 \AA\ intervals starting at 920 \AA\, are presented in Table~\ref{tab2}. 

We can estimate the amount of contamination that the OTA has suffered by comparing a model of the effective area to these data.  Recall that the effective area is the product of the telescope geometric area, the reflectivity of each telescope mirror, the grating absolute efficiency and the detector efficiency.  We use the measured MgF$_2$/Al reflectance (250 \AA\ of MgF$_2$) from \citet{Hunter:1971} for the mirrors.   G140L absolute efficiency measurements were not available for the full far-UV wavelength range, so we used those available from \citet{Osterman:2002} to constrain a model computed as the product of the MgF$_2$/Al reflectance and a groove efficiency for a laminar profile \citep{McCandliss:2001}.  The COS detector QEs measured by \citet{Vallerga:2001} were further augmented with additional CsI photocathode measurements at the short wavelengths from similar detectors by the same group.  

The dashed line on Figure~\ref{fig2} is the resulting model for the effective area.  Multiplying our model by a factor $\sim$ 0.6 produces a reasonable match to the G140L Segment B effective area. If we assume that the OTA carries the bulk of the disagreement and that the original reflectivity was similar to that of \citet{Hunter:1971}, then after accounting for the two reflections, we conclude the individual mirrors of the OTA retain $\approx$ 80~\% of their original efficiency. 


\section{Discussion}

The effective area of COS at wavelengths below 1000 \AA\ is comparable to the effective area of {\it FUSE} over the same bandpass and it exceeds that of {\it FUSE} in the 1090 -- 1180 bandpass \citep{Kruk:2009}.  At present, the error bars at the short wavelength end are somewhat uncertain because of the monotonically decreasing count rate in our calibration object, caused by the overlap in the converging pressure broadened Lyman series, but this situation will improve as further calibrations are acquired.

The G140L has a resolving power of $R \sim$ 2000, which is $\sim$ 10 times lower than than the {\it FUSE} resolution.  As such, the detector background limit of the G140L segment B should be correspondingly lower provided the detector dark rate is similar to that realized by {\it FUSE}.  The combination of reasonable effective area with low detector background will allow COS to pursue science left behind by {\it FUSE} when the planned low spectral resolution channel was descoped.   We briefly discuss some of these science opportunities here.

For example, analysis of  \ion{He}{2} ``Gunn-Peterson'' absorption in the intergalactic medium (IGM) carried out by {\it FUSE} was conducted primarily on two quasar lines-of-sight HE 2347-4342 and HS 1700+6416 \citep{Kriss:2001, Shull:2004, Zheng:2004, Fechner:2006}.  The {\it FUSE} observations in combination with \ion{H}{1} absorption line studies from the ground demonstrated that fine-grained fluctuations in the metagalactic ionizing background are associated with source hardness, IGM density voids and peaks. However, there is a critical need for multiple lines-of-sight to better constrain the cosmic variance in the evolution of these peaks and voids and in the redshift associated with the initiation of the \ion{He}{2} reionization epoch near $z_r$(\ion{He}{2}) = 2.8 $\pm$ 0.2 \citep{Shull:2006}.

The higher sensitivity at low flux will also be beneficial to programs seeking to detect and quantify the fraction of Lyman continuum photons that escape from  star-forming galaxies at low redshift.  Low redshift observations are essential \citep{McCandliss:2008} to further our understanding of the physical processes responsible for the \ion{H}{1} reionization epoch near $z_r$(\ion{H}{1}) = 6 thought to be caused by Lyman continuum photons escaping from the first star-forming galaxies \citep{Yan:2004}.

COS will be probing clouds in the interstellar medium with total extinction $A_V > $ 3  where the long sought after transition from translucent to dense medium is expected to occur \citep{Rachford:2009}.  One of the problems that these observing programs faced with the demise of {\it FUSE} and the uncertain COS far-UV response was the lack of guaranteed access to the dominate cloud species, H$_2$, which has ground state absorption bands in the far-UV shortward of $\sim$ 1110 \AA.  This is no longer a problem now that the COS G140L Segment B has been shown to have useful sensitivity. It will now be possible, for the first time, to make direct studies of the CO/H$_2$ ratio \citep{Burgh:2007} in dense regimes more typical of those probed by CO observations at radio wavelengths.  Ultimately this could lead to more accurate estimations of molecular gas mass in star-forming galaxies.

In addition to the enabled science investigations discussed above, spectral coverage below 1050 \AA\ provides unique access to the \ion{O}{6} $\lambda\lambda$~1032, 1038 resonance doublet.  This transition is perhaps the best tracer of gas at $T$~$\approx$~few~$\times$~10$^{5}$ K in the UV-optical bandpass and can be used to constrain physical processes in a diverse range of galactic and intergalactic environments.

\section{Conclusion}

We have demonstrated that there is significant response in the G140L Segment B channel of COS shortwards of the MgF$_2$/Al discontinuity near 1100 \AA\  extending all the way to the Lyman edge.  This low resolution far-UV capability is a unique instrumental mode that has never before been available to the astronomical community and as such has great potential for scientific discovery.

The COS far-UV demonstration has implications for future facility class missions. We note that no effort was made to optimize the far-UV response of this channel. The use of a SiC coated grating would provide a factor of two gain in efficiency.  An additional factor of two would result through the use of a dual order spectrograph design \citep{Lupu:2008} or a blazed ion etched grating.  We estimate that the efficiency of a properly optimized spectrograph on a general purpose 10 meter telescope, having a two bounce OTA with standard MgF$_2$/Al mirrors, has the potential to achieve an effective area $\sim$ 500 to 1000 cm$^{2}$ in the far-UV, easily breaking the desired factor-of-10 threshold commonly used to assess the discovery potential of new observing modes \citep{Moos:2004}.  The COS observations presented here prove that such an approach is a viable option for large facility class missions and suggests that including a windowless far-UV channel below 1150 \AA\ on future missions should be seriously considered.

\acknowledgments

The authors would like to acknowledge useful and encouraging discussion with the  following people: Cynthia S. Froning, J. Michael Shull, John T. Stocke, Theodore P. Snow, S. Alan Stern, Derck L. Massa, Oswald H. W. Siegmund, John V. Vallerga, Brian Fleming, Roxana E. Lupu, Eric B. Burgh, David J. Sahnow, Jeffrey W. Kruk, W. Van Dixon, William P. Blair, Kenneth Sembach, B-G Andersson, Paul D. Feldman, and H. Warren Moos.  This work has been supported by NASA contract NAS5-98043 to the University of Colorado.



{\it Facilities:}  \facility{HST (COS)}.

\clearpage



\clearpage

\begin{table}
\begin{center}
\caption{Spectral data files \label{tab1}}
\begin{tabular}{lr}
\tableline\tableline
Instrument & Spectral File \\
\tableline
COS 		& la9h02nsq\_corrtag\_b.fits \\
\ldots 		& la9h02nsu\_corrtag\_b.fits \\
\ldots 		& la9h02nsz\_corrtag\_b.fits \\
{\it FUSE} 	& D02302011031bsic4ttagfcal.fit \\
\ldots		& D02302011031alif4ttagfcal.fit \\
\ldots 		& D02302011032alif4ttagfcal.fit \\
\tableline
\end{tabular}
\end{center}
\end{table}

\begin{table}
\begin{center}
\caption{G140L Segment B Effective Area and Errors\label{tab2}}
\begin{tabular}{rrr}
\tableline\tableline
$\lambda$ (\AA) & A$_{eff}$ (cm$^2$) & $\pm$ 1$\sigma$\\
\tableline
     920 &    10. &     6. \\
     940 &    11. &     3. \\
     960 &    12. &     2. \\
     980 &    10. &     2. \\
    1000 &     9. &     2. \\
    1020 &    11. &     2. \\
    1040 &    19. &     2. \\
    1060 &    19. &     2. \\
    1080 &    35. &     3. \\
    1100 &   215. &    17. \\
    1120 &   654. &    33. \\
    1140 &  1133. &    35. \\
    1160 &  1377. &    43. \\
\tableline
\end{tabular}
\end{center}
\end{table}





\end{document}